# Ultra-low phase-noise microwave generation using a diode-pumped solid-state laser based frequency comb and a polarization-maintaining pulse interleaver

**Erwin Portuondo-Campa, Gilles Buchs, Stefan Kundermann, Laurent Balet and Steve Lecomte[*]**

*Centre Suisse d'Electronique et de Microtechnique (CSEM), Jaquet-Droz 1, 2000 Neuchâtel, Switzerland*
[*]*steve.lecomte@csem.ch*

**Abstract:** We report ultra-low phase-noise microwave generation at a 9.6 GHz carrier frequency from optical frequency combs based on diode-pumped solid-state lasers emitting at telecom wavelength and referenced to a common cavity-stabilized continuous-wave laser. Using a novel fibered polarization-maintaining pulse interleaver, a single-oscillator phase-noise floor of -171 dBc/Hz has been measured with commercial PIN InGaAs photodiodes, constituting a record for this type of detector. Also, a direct optical measurement of the stabilized frequency combs timing jitter was performed using a balanced optical cross correlator, allowing for an identification of the origin of the current phase-noise limitations in the system.

**1. Introduction**

In recent years, photonics-based technologies have opened new routes for the generation of ultra-low phase-noise microwave signals. Using an optical frequency comb (OFC) as a coherent frequency divider allows transferring the ultra-high stability of a continuous-wave (CW) laser locked to a reference cavity to the microwave domain [1],[2],[3]. This approach has been implemented with different types of femtosecond lasers, including Ti:sapphire [3] and Er-doped fiber oscillators [2], [4]. Ultra-pure microwave generation from an OFC based on diode-pumped solid state lasers (DPSSL) has also been demonstrated for Yb:KYW gain medium [5] and initial trials have been conducted with a DPSSL based on Er:Yb:glass [6]. This type of laser has shown excellent timing stability with sub-100 attosecond integrated timing jitter (from 10 kHz to 50 MHz) and very low amplitude noise [7]. Furthermore, it is particularly suited for operation in challenging environments as it has been recently reported in the context of space applications [8].

The photodetector providing the optical to electrical conversion of the OFC pulse train is a critical element in the photonic microwave generation scheme. Especially at offset frequencies far from the carrier, photodetection shot noise imposes a fundamental limit to the microwave signal phase-noise which can be orders of magnitude higher than the intrinsic noise of the optical oscillator. The situation can be improved with a high average direct photocurrent together with high repetition rates close to the desired microwave frequency in order to increase the photodiode saturation limit [9]. Furthermore, recent works have shown that shot noise correlation effects in the photodetection of short optical pulse trains can significantly reduce phase-noise below continuous wave photodetection noise floors [10],[11]. However, photocarrier scattering and distributed absorption set limitations to this effect [12].

Standard microwave oscillators for a broad set of applications have typical frequencies in the range of 10 GHz. Mode-locked lasers on the other hand typically operate in stable regimes for pulse repetition rates up to a few hundreds of MHz. Thus, a high harmonic of the photo-detected signal is generally used as microwave output, whose power can be increased through multiplication of the optical repetition rate by using a cascaded fibered pulse interleaver [4], [13].

Here, we demonstrate record low far-from-carrier phase-noise for photonically generated microwave signals from passively mode-locked DPSSLs based on Er:Yb:glass gain medium coupled to polarization-maintaining low-loss fiber pulse interleavers and using commercial fast PIN InGaAs photodetectors. In addition, an optical measurement of the pulse train timing jitter allowing for an identification of the origin of the phase-noise limitations in the current system is presented.

## 2. Experimental setup

Two OFCs identical by design were used to produce two ultra-pure microwave sources which could be compared for characterization of their phase-noise. A diagram of the experimental setup is presented in Fig. 1.

The OFCs were generated by DPSSLs based on Er:Yb:glass gain medium. The lasers design followed a typical architecture as described in detail in Ref. [14], with a semiconductor saturable absorber mirror (SESAM) for passive mode-locking. The lasers emit transform-limited 146 fs soliton pulses at 1556 nm with average powers of about 150 mW at a 100 MHz pulse repetition rate.

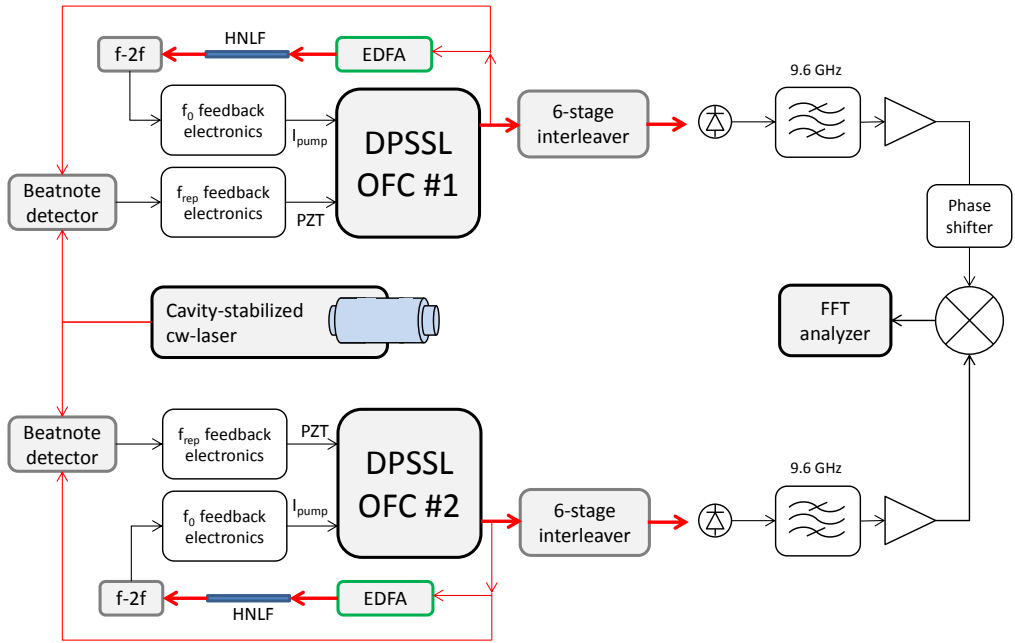

Fig. 1 . General diagram of the experimental setup. Red arrows indicate light transmission paths. Black arrows correspond to electronic signal paths. DPSSL: diode-pumped solid-state laser, OFC: optical frequency comb, EDFA: Erbium-doped fiber amplifier, HNLF: highly nonlinear fiber, PZT: piezoelectric transducer, FFT: fast Fourier transform.

Approximately 75% of the optical power generated by each OFC was fed into 6-stage pulse interleavers multiplying the initial 100 MHz repetition rate to 3.2 GHz (the third harmonic at 9.6 GHz is used). The remaining 25% of the optical power, representing about 40 mW were further split with a 50/50 fiber splitter and used for the stabilization of the OFC repetition rate ($f_{rep}$) and offset ($f_0$) frequencies.

The part of the optical signal used for the detection of $f_0$ was initially amplified in an Erbium-doped fiber amplifier (EDFA) and injected into a highly nonlinear fiber (HNLF) for generation of an octave-spanning spectrum signal feeding an f-2f interferometer. The feedback electronics of $f_0$ follow a similar scheme as presented in Ref. [15], with the correction signal applied to the pump diode injection current ($I_{pump}$). The RF references used to phase-lock the $f_0$ signals were generated by two independent synthesizers, set at the same frequency and both referenced to a common active hydrogen maser (not shown in Fig. 1)

The part of the optical signal used for the stabilization of $f_{rep}$ was filtered with a narrow optical band-pass filter and combined with an ultra-stable cw-laser. The resulting heterodyne beatnote was used for the stabilization of $f_{rep}$ using a phase lock loop scheme as described in Ref. [5], with the correction signal applied to a piezo-electric transducer (PZT) mounted in the DPSSL cavity. The cw-laser was a commercial distributed feedback fiber laser, stabilized to a compact high-finesse Fabry-Perot cavity (4 cm long) through a home-made Pound-Drever-Hall setup, providing Hz-level optical linewidth at 1 sec time scale and typical long-term drifts below 100 mHz/s. Two independent synthesizers set at the same frequency and referenced to the same active hydrogen maser were used as references to phase-lock the beat note signals. Since only one single cw-laser served as a reference for locking $f_{rep}$ for both OFCs, the measured phase-noise of the microwave signals was insensitive to any drifts of the cw-laser frequency, as it was insensitive to the drift of the hydrogen maser. As a result, microwave phase-noise in the locking bandwidth of $f_{rep}$ is relative, whereas it is absolute for offset frequencies above the locking bandwidth cut-off.

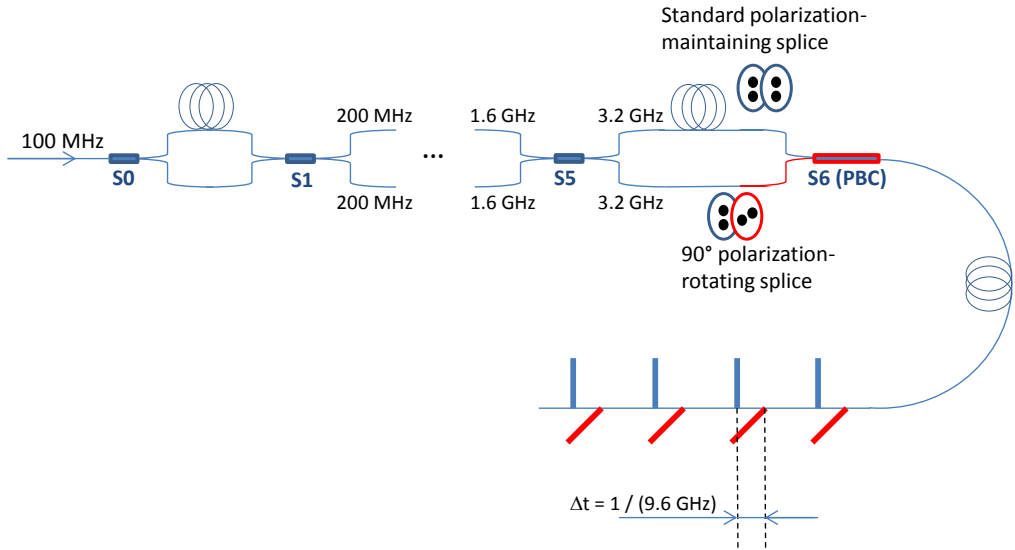

Fig. 2. Interleaver design. S0 to S5 correspond to PM splitters/combiners which constitute the 5 first interleaver stages. The last stage is ended by a fibered polarizing beam combiner: S6 (PBC). Red and blue colors indicate orthogonal polarizations.

The design of the pulse interleavers is presented in Fig. 2. For each pulse interleaver, five consecutive stages made of commercial PM 50/50 fiber splitters (split ratio error < 1%) were first assembled by fusion splicing and designed to multiply the pulse repetition rate of the DPSSL by a factor $2^5$, i.e. from 100 MHz to 3.2 GHz. At this point, it is possible to generate a microwave signal close to 10 GHz by taking the third harmonic of the photodetected pulse train at 9.6 GHz. In such approach however, the harmonics at 3.2 GHz and 6.4 GHz would generally present larger amplitudes than the component at 9.6 GHz, thus reducing the effective saturation power of the useful microwave signal. This drawback was remediated with an additional 6$^{th}$ stage where the delay introduced between the two arms is an entire multiple of the period of the useful signal [9]. In our case a delay of 104.2 ps = (9.6 GHz)$^{-1}$ was added. One advantageous feature implemented in our pulse interleavers was the use of a fiber polarizing beam combiner at the final stage. A 90° rotated splice (slow axis spliced on fast axis) on one input arm as shown in Fig. 2 allowed recovering all the optical power at a single output. Since the photodiodes are polarization insensitive, such a construction allows for a net output power almost 3 dB higher than in the case of non-polarizing fiber combiners with two outputs. Using a fast sampling scope with a nominal time resolution of 0.2 ps, the delay error on all six stages could be maintained ≤ 1.5 ps.

Two highly linear commercial PIN InGaAs photodiodes with respective bandwidths of 12 GHz and 23 GHz (Discovery Semiconductors HLPD DSC50S and HLPD DSC30S, respectively) were used to detect the optical signals. The incident optical power on the photodiodes could be adjusted via the optical coupling into the fiber interleavers. The produced electrical signals were band-pass filtered to isolate the 9.6 GHz microwave tone and further amplified (power level adjusted with attenuators for higher photocurrent settings) to feed a double balanced mixer in phase-detector configuration and operating in a regime where the AM sensitivity is minimized. A phase shifter in one of the microwave paths ensured reaching the quadrature condition for phase-noise measurements. The power spectral density (PSD) of the mixer IF output was then measured with an FFT analyzer and converted into phase-noise with independent calibration of the mixer for each set of LO and RF powers used in the experiments.

The intrinsic timing jitter of the optical pulse trains at 100 MHz was measured independently from the microwave setup using a balanced optical cross-correlator (BOC). Previous measurements were performed with quasi-free running lasers (300-Hz-bandwidth synchronization of repetition rates only) [7]. In this work, the BOC measurements have been performed with fully locked OFCs operated in the regime used for microwave generation. In contrast to Refs. [7] and [16], here no feedback from the BOC signal is necessary to synchronize the pulse trains since the repetition rates of the OFCs are locked to the same optical reference. The BOC measurement allows for a direct comparison between the optically-generated phase-noise and the final microwave phase-noise.

## 3. Results

*3.1 OFC stabilization and relative intensity noise*

The stabilization of the OFCs was achieved with lock bandwidths close to 40 kHz for both $f_{rep}$ and $f_0$ loops. Fig. 3 shows typical spectra of the stabilized beatnote signals $f_0$ and optical beatnote (for $f_{rep}$), as well as the typical relative intensity noise (RIN) spectrum of the DPSSLs (Fig. 3 (c)) for the conditions of free-running laser, $f_0$-only stabilization and full OFC stabilization ($f_0$ and $f_{rep}$). It can be observed that the stabilization of $f_0$ significantly reduces the laser RIN within the lock bandwidth. This is a well-known effect for DPSSLs resulting from reduced pump laser power fluctuations producing in turn lower intracavity and output power fluctuations [15]. On the other hand, the additional stabilization of $f_{rep}$ has no sensible effect on the laser RIN.

*3.2 Pulse interleaver outputs and RF spectrum of photo-detected signals*

The losses within the interleavers (fiber splices and losses at the combiners) were estimated to 1 – 1.5 dB, with the rest of the missing optical power being lost at the fiber coupling. It is worth noticing that without the PM configuration and the polarizing beam combiner on the last interleaver stage, the available power out of each interleaver (up to 67 mW) would be halved and the maximum admissible photodiode current could not be reached. Fig. 4 shows typical microwave spectra taken at the output of the 12 GHz-bandwidth photodiode with 5 (blue dots) and 6 (red curve) stage interleavers. In both cases the optical power was adjusted to produce an average photocurrent of 11 mA. The addition of the 6[th] interleaver stage with tailored inter-pulse delay allowed to increase the microwave power in the 9.6 GHz tone, now exceeding the 3.2 GHz and 6.4 GHz tones by roughly 3 dB. Both interleavers produced similar RF spectra. After band-pass filtering of the microwave signal, the power in the closest peaks at ± 100 MHz from the 9.6 GHz tone was at least 28 dB below the carrier for both oscillators. Fig. 5 shows the photodiode microwave output power (corrected for cable losses) at 9.6 GHz as a function of the photocurrent. The optical power coupled into the pulse interleavers was adjusted to produce the same average photocurrent in both photodiodes. The latter were set at a bias voltage of 9 V and temperature stabilized to 20°C.

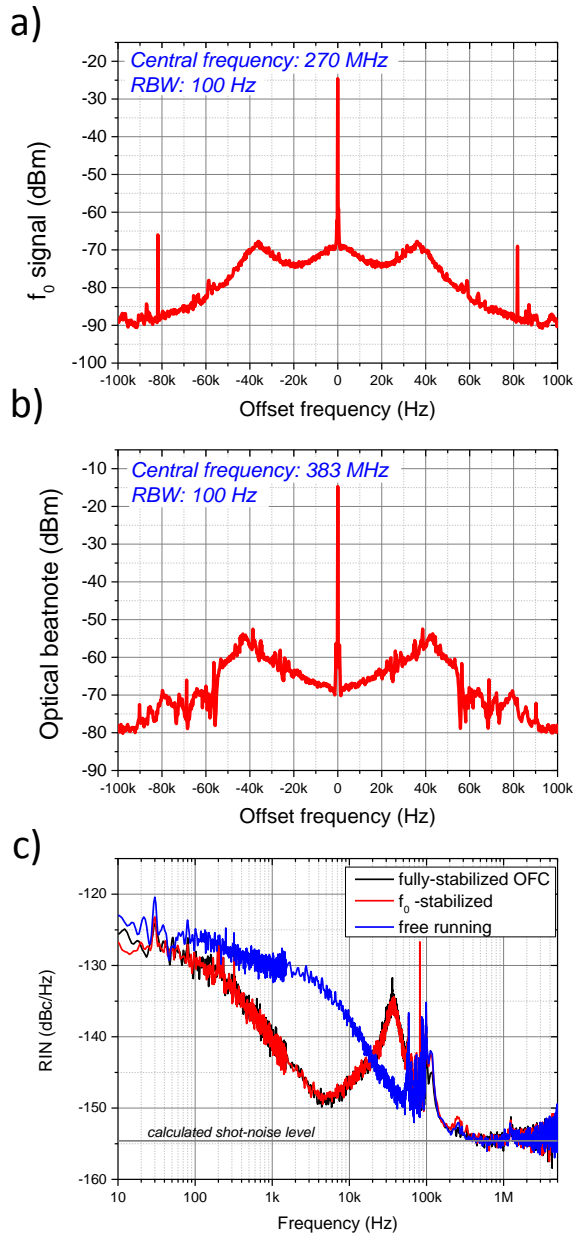

Fig. 3. a) and b) RF spectra of the $f_0$ and optical beat-note signals (for $f_{rep}$) of a stabilized OFC. c) RIN spectrum of one of the DPSSL in the conditions: free-running (blue curve), $f_0$ – stabilized (red curve) and fully stabilized (black curve) as a frequency comb.

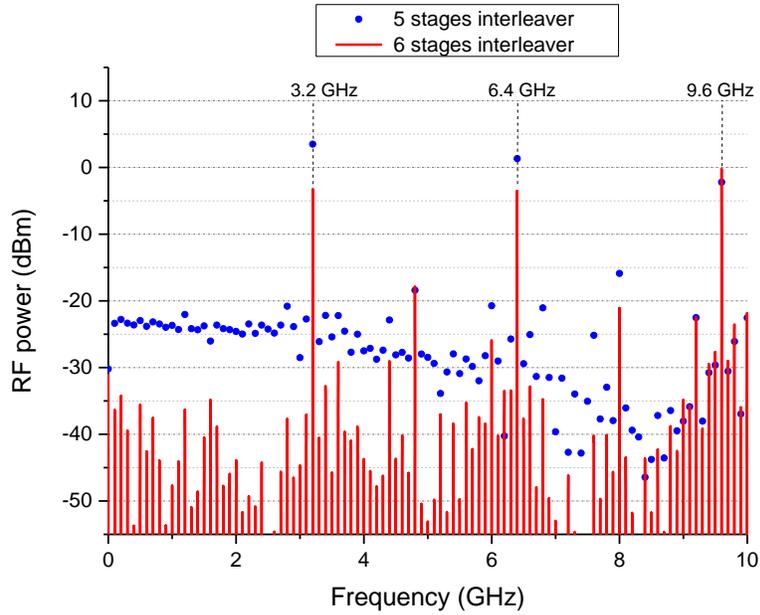

Fig. 4. Microwave spectra of a photodiode signal without filtering or amplification as obtained with a 5-stages pulse interleaver (blue dots) and with a 6-stages pulse interleaver (red lines) for the same average photocurrent.

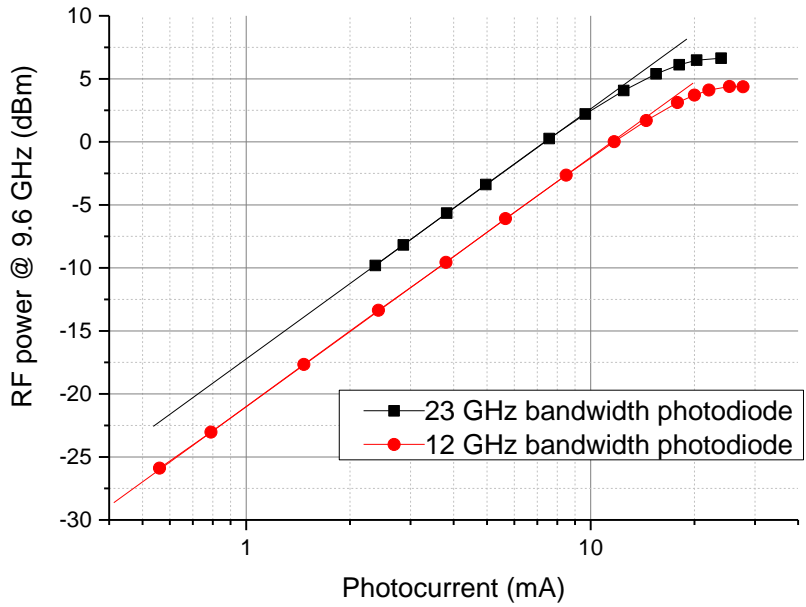

Fig. 5. Photodiodes (23 GHz and 12 GHz bandwidths) microwave output power at 9.6 GHz as a function of the photocurrent.

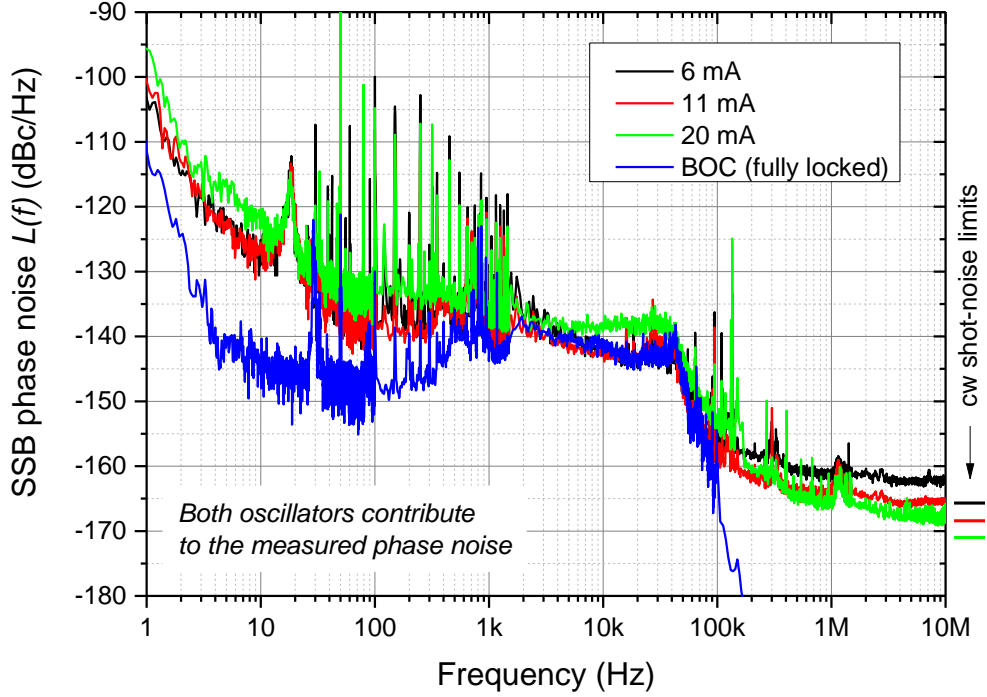

Fig. 6. Single side-band phase-noise $\mathcal{L}(f)$ of the photo-generated 9.6 GHz microwave carriers measured relative to each other for different photocurrents (black, red and green curves). The blue curve corresponds to the balanced optical cross correlator (BOC) measurement of the stabilized lasers, scaled to the carrier frequency of 9.6 GHz. For comparison, calculated continuous-wave shot-noise limits are indicated on the right of the plot.

**Table 1. Calculated cw shot-noise limits for both oscillators (according to Ref. [11]), measured phase-noise floors and excess measured phase-noise compared to the cw shot-noise limit for three different photocurrent values.**

| Photocurrent (mA) | Calculated cw shot-noise limit (dBc/Hz) | Measured phase-noise floor (dBc/Hz) | Excess measured phase-noise (dB) |
|---|---|---|---|
| 6 | -165.7 | -162.4 | 3.3 |
| 11 | -168.4 | -165.4 | 3 |
| 20 | -171 | -168 | 3 |

*3.3 Microwaves phase-noise and stabilized OFC timing jitter*

Single side-band phase-noise spectra $\mathcal{L}(f)$ containing noise contributions from both 9.6 GHz microwave sources are displayed in Fig. 6 for different photocurrent values. The measured spectrum of optical timing jitter of the fully stabilized OFCs was converted to the equivalent phase-noise for a carrier frequency of 9.6 GHz and is presented on the same graph. From these plots, three regions of interest can be distinguished:

a) 1 Hz – 1 kHz: The microwave phase-noise is not limited by the laser timing jitter, which shows extremely low values down to -148 dBc/Hz (both oscillators contributing). The same interval contains a number of spurs, mainly harmonics of 50 Hz attributed to technical noise. In the 1-Hz offset range, the phase-noise is in the order of - 100 dBc/Hz, which is typical for this kind of oscillator [2],[3],[17]. The optical reference being common to both oscillators, its own frequency stability does not affect the measured phase-noise. Therefore in our system the close-to-carrier phase-noise must be attributed to phenomena outside the OFC stabilization loops, such as mechanical vibrations in the interleaver fibers or within the free-space optical path to couple the laser into the interleaver. As will be discussed later, in the 30 Hz to 200 Hz frequency range, the phase-noise is essentially limited by the amplifier residual phase-noise (see Fig. 7).

b) 1 kHz – 100 kHz: The laser timing jitter limits the microwave phase-noise for 6 mA and 11 mA photocurrents. For 20 mA, the phase-noise is slightly degraded. Since the optical jitter of the laser pulses is intrinsically the same for all photocurrents, the most plausible mechanism that can explain the observed power dependence is AM-to-PM conversion at the photodetection stage. It is known that the AM-to-PM conversion coefficient of a photodiode generally depends on the optical power and can reach 0-value at certain photocurrent settings, depending on the photodiode design and even changing from one device to another within the same model [18], [19]. Here, these coefficients have not been determined, but our results suggest that AM-to-PM conversion tends to increase with photocurrent and diode saturation at least up to 20 mA photocurrent.

c) 100 kHz – 10 MHz: The laser intrinsic optical phase-noise is much lower than the measured phase-noise of the microwave signals. At an offset frequency of 10 MHz, all the phase-noise curves have reached a plateau with their values reported in Table 1. For a photocurrent of 20 mA, the combined noise floor reaching -168 dBc/Hz (-171 dBc/Hz for a single oscillator assuming identical sources) represents to the best of our knowledge the lowest phase-noise floor ever demonstrated for photonics-generated microwaves using InGaAs PIN photodetectors [9], [13].

To set the context for the values of the measured phase-noise plateau (at 10 MHz offset frequency), the continuous-wave (cw) shot-noise limit for both oscillators has been calculated [11] for each photocurrent value without considering shot-noise correlations linked to short pulses photodetection described in Ref. [10]. The calculated values, reported in Table 1 and displayed next to the measured phase-noise spectra in Fig. 6 show that all three plateaus exceed the corresponding calculated cw shot-noise limits by about 3 dB. To better understand the nature of this excess noise, the experimental noise corresponding to a photocurrent of 11 mA has been determined for different conditions. For this, two microwave signals at 9.6 GHz, split from a single synthesizer, were feeding the mixer with the microwave powers adjusted in order to match the levels used in the 11 mA photocurrent microwave phase-noise measurement. The phase shifter was tuned to reach the quadrature condition and a phase-noise acquisition was performed with all the same instrumental settings as for the photonics-generated microwaves, in a first stage without the microwave amplifiers. The resulting phase-noise spectrum is shown in Fig. 7 (black curve). A residual phase-noise 10 dB below the 11 mA noise floor is obtained at 10 MHz offset frequency. In a second stage, the measurement was repeated including the amplifiers in the microwave paths and readjusting the powers conveniently. The resulting phase-noise spectrum (dark green curve) clearly shows that the residual phase-noise of the amplifiers limits the achieved phase-noise floor far from the carrier. The same conclusion is valid for the 6 mA and 20 mA photocurrent cases. Notice that the dependence of the mixer response with the input powers affecting the phase-discriminator value also influences slightly the resulting instrumental noise floor. In addition, the residual phase-noise of the amplifiers has been measured individually as well and confirms this limit. The use of photodetectors with higher saturation photocurrent such as modified uni-traveling carrier (mUTC) photodiodes [20] should allow reaching sufficient microwave powers for driving the mixer without amplifiers, thus allowing the observation of correlation induced sub-cw phase-noise floors as reported in Refs [10], [11] for short-pulses photodetection. State-of-the-art modified uni-travelling carrier

(mUTC) photodiodes can work in the linear regime for optical powers above 100 mW and can generate 10 GHz carrier signals above 10 dBm without amplification [21].

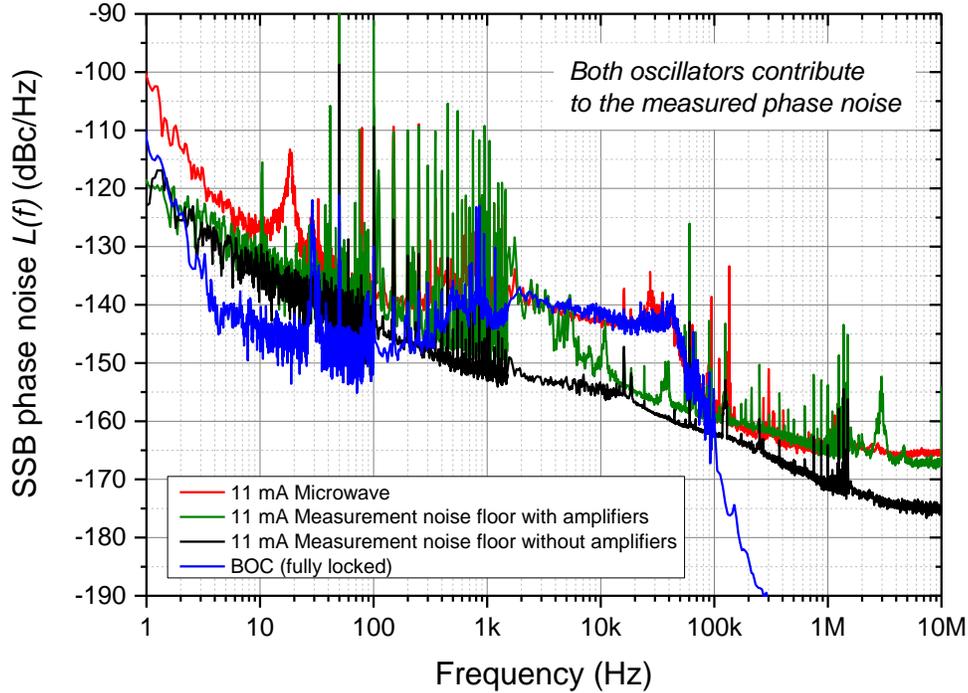

Fig. 7. Single side-band phase-noise $\mathcal{L}(f)$ of the photo-generated 9.6 GHz microwave carriers measured relative to each other for 11 mA photocurrent (red curve), balanced optical cross correlator (BOC) measurement (blue curve), measurement noise floor with the microwave amplifiers (dark green curve) and measurement noise floor without amplifiers (black curve).

## 4. Summary and outlook

Ultralow phase-noise microwave generation through the photodetection of a stabilized passively mode-locked DPSSL has been demonstrated using commercially available InGaAs PIN photodetectors. In order to increase the available power of a carrier microwave near 10 GHz, the laser pulses were multiplied with polarization-maintaining fibered pulse interleavers featuring five stages of repetition rate multiplication ($2^5$-factor) plus one final stage designed to enhance the desired harmonics at 9.6 GHz. In addition, the final stage conveniently combined the total power from both interleaver arms into a single output channel using a polarizing beam combiner. With this strategy, it was possible to produce the lowest far-from-the-carrier phase-noise floor so far using PIN photodetectors (-171 dBc/Hz @ 10 MHz offset, single source contribution). This phase-noise floor is currently limited by the residual phase-noise of the microwave amplifiers and prevents observation of shot-noise correlations leading to phase-noise floors below the cw shot-noise limit. Finally, an optical measurement of the timing jitter of the laser pulse trains for fully stabilized OFCs provided the intrinsic phase-noise of the systems without the interleavers and the photodetection stages. This measurement confirmed that the current architecture (PIN photodiode and microwave amplifiers) limits the phase-noise floor far from the carrier. Combining the high stability of the optical oscillator described here with the large power-handling capabilities of mUTC photodetectors [21] should allow in the near future reaching even lower phase-noise performances for demanding microwave applications.

**Acknowledgments**

The authors acknowledge J. Bennès for the electronics hardware, as well as F. Quinlan and S. A. Diddams for fruitful discussions. This work was funded by the FP7-SPACE-2013-1-607087-PHASER project and the Canton of Neuchâtel.